\documentstyle[aps,manuscript,epsf]{revtex}

\font\bgreek=cmmib10

\def\Btheta{\hbox{\bgreek\char018}}

\def\Btau{\hbox{\bgreek\char028}}

\def\RI{{\rm I}}

\def\r{{\bf r} }

\def\q{{\bf q} }
\def\A{{\bf A}}
\def\B{{\bf B}}
\def\H{{\bf H}}
\def\j{{\bf j}}
\def\q{{\bf q}}

\def\tr{{\rm tr}}
\def\Bnabla{{\bf \nabla}}
\def\cp{{\cal P}}
\font\msxm=msxm10  
\def\W{{\bf W}} \def\cl{{\cal L}} 
\def\ch{{\cal H}} \def\cd{{\cal D}}

\def\got{\,\hbox{\msxm \char38}\,}
\def\la{{\langle}}
\def\ra{{\rangle}}
\title{Polymer Winding Numbers and Quantum Mechanics$^*$}
\author{David R. Nelson$^{(1)}$, \\
and Ady Stern$^{(2)}$}
\address{$^{(1)}$Lyman Laboratory of Physics, 
Harvard University,
Cambridge, MA 02138}
\address{$^{(2)}$Department 
of Condensed Matter Physics,
Weizmann Institute of Sciences,
Rehovot 76100, Israel}
\date{\today}
\begin{document}
\draft
\maketitle
\widetext
\begin{abstract}
The winding of a single polymer in thermal equilibrium around 
a repulsive cylindrical obstacle is perhaps the simplest 
example of statistical mechanics in a multiply connected
geometry. 
As shown by S.F. Edwards, this problem
is closely related to the quantum mechanics of a
charged particle interacting with a Aharonov-Bohm
flux. In another development, Pollock and
Ceperley have shown that boson world lines in
$2+1$ dimensions with periodic boundary
conditions, regarded as ring polymers on a torus,
have a mean square winding number given by
$\langle W^2\rangle=2n_s\hbar^2/mk_BT$, where $m$ is
the boson mass and $n_s$ is the superfluid number
density. Here, we review the mapping of the
statistical mechanics of polymers with constraints
onto quantum mechanics, and show that there is an
interesting generalization of the
Pollock-Ceperley result to directed polymer melts
interacting with a repulsive rod of radius $a$.
When 
translated into boson language, the mean square winding 
number around the rod for a
system of size $R$ perpendicular to the rod reads
$\la W^2\ra={n_s\hbar^2\over 2\pi mk_BT}\ln(R/a)$.
This result is directly applicable to vortices in
Type II superconductors in the presence of
columnar defects. An external current passing
through the rod couples directly to the winding
number in this case.
\bigskip

\end{abstract}
\narrowtext
\section{INTRODUCTION}

The study of the statistical mechanics of polymers
in multiply connected geometries began many years
ago with work by S.F. Edwards and by S. Prager and
H.L. Frisch \cite{edwards,prager}. The simplest nontrivial
geometry consists of a polymer interacting with a
repulsive rod, and the corresponding path
integrals can be analyzed via an elegant analogy
with the physics of a quantum mechanical particle
interacting with a solenoidal vector potential
\cite{gross}. The mean square winding number of the polymer
around the rod, and other interesting quantities
can be computed for this problem \cite{rudnick}. The physics
bears a close mathematical relation to the famous
Aharonov-Bohm effect for a real quantum
mechanical particle interacting via its charge
with a tube of magnetic flux \cite{aharonov}. 

More recently, Pollock and Ceperley have studied
the winding numbers with respect to periodic
boundary conditions for boson world lines in the
Feynman path integral formulation of superfluidity
in two dimensions \cite{pollack}. The physics here is
equivalent to {\it many} ring polymers
interacting on the surface of a torus. The mean
square winding number of the world lines around
the torus can be expressed exactly in terms of the
renormalized superfluid density of the equivalent
boson system \cite{ceper}.

Winding numbers and the statistical mechanics of
many polymer-like objects are also relevant to the
physics of vortex lines in Type II superconductors
\cite{blatter}. Here, thermal fluctuations in the trajectory
of a vortex defect in the superconducting order
parameter can be described by a Feynman path
integral for an elastic string. A collection of
many such lines behaves like a directed polymer
melt, with the added complication of a quenched
random disorder potential \cite{nelson}. Point-like disorder
is always present to some degree due, for example,
to quenched fluctuations in the concentration of oxygen
vacancies in the high temperature cuprate
superconductors. Drossel and Kardar have studied
how the winding number  distribution of a single
directed polymer around various obstacles is
affected by point-like disorder \cite{drossel}.

Dramatic improvements in the pinning efficiency of
vortex lines have recently been achieved via the
introduction of {\it columnar} damage tracks
created by heavy ion irradiation. If the
concentration of damage tracks (assumed to pass
completely through the sample) exceeds the number
flux lines, there is a low-temperature ``Bose
glass'' phase, in which every vortex is trapped on
a columnar defect \cite{nelson93}. At high temperatures,
the vortices delocalize in an entangled flux
liquid. If these vortex trajectories are viewed as
the world lines of quantum mechanical particles,
the physics in thick samples becomes equivalent to
that of a low temperature boson superfluid in
$2+1$ dimensions \cite{nelson}.

In this paper, we discuss how to compute polymer
winding numbers from the Hamiltonian formulation
of quantum mechanics, keeping in mind applications
to vortex lines interacting with many columnar
defects. A random distribution of columns maps
onto a time-independent random potential in the
quantum mechanical analogy \cite{nelson93}. We assume high
temperatures, and samples which are clean
before irradiation, so
that point disorder is negligible and the usual
Abrikosov flux lattice is either melted by thermal
fluctuations or destroyed by the columnar
disorder. An experimental realization of a
multiply connected geometry for thermally excited
flux lines is illustrated in Fig.~1. Superimposed
on a dilute concentration of parallel columnar
defects scattered randomly throughout a
superconducting sample is a slender tube in which
the columns are very dense. Such a ``tube of
columns'' could be made by covering a sample with
a mask containing a small submicron hole during
irradiation with a strong dose of heavy ions.
Imagine that this sample is first subjected to a
very large magnetic field, such that the density
of flux lines is approximately equal to the
density of columns in the tube. It should be
possible to choose the temperature such that the
vortices in the tube are in the Bose glass phase,
while those outside constitute a flux liquid. The
many trapped vortices inside the tube should then 
present a virtually impenetrable barrier to the
thermally excited vortex lines outside. The
concentration of lines in the liquid outside this 
repulsive cylindrical obstacle could
be varied by  decreasing the magnetic field.
Because flux creep in the Bose glass phase is
quite slow \cite{blatter,nelson93}, the concentration of vortices in
the tube should remain approximately constant as a
field is turned down, leaving the barrier almost
unchanged. 

We shall explore the winding number fluctuations
of the vortices with respect to the repulsive
cylindrical tube in this experiment. As discussed
below, a current passing through the tube couples
directly to the net vortex winding number, and the
mean square winding number of the unperturbed
system gives the linear response of the net
winding number to this current. (See 
Fig.~2) Such a current
acts like an imaginary vector potential when this
problem is mapped onto quantum mechanics \cite{nelson93}.
Winding of vortices around a thin repulsive
obstacle (or a set of such obstacles) could be
probed via double-sided flux decorations
\cite{yao}  or indirectly by monitoring the flux flow resistivity in
the plane perpendicular to the common direction of
the applied field and the columns. The
extra entanglements induced by the longitudinal current
should impede vortex transport in this plane, and
it would be especially interesting to look for
changes in the in-plane resistivity as a function
of the current through the tube \cite{text}. This
resistance should drop with increasing
longitudinal current through the tubes, due to
the enhanced vortex winding about the obstacle. 

In section II, we briefly review the Feynman path
integral description of bosons in $2+1$
dimensions, and discuss the Pollack-Ceperley
result for the winding numbers of bosons on a torus. 
We then describe the closely related
statistical mechanics of directed polymer melts
and vortex lines, and show how a current through
a cylindrical obstacle couples to the winding
number. In section III, we treat winding number
statistics for isolated  polymers both with and
without columnar pin disorder. Results for {\it
many} polymers or flux lines winding about an
obstacle are presented in section IV. 

\section{TWO-DIMENSIONAL BOSONS AND DIRECTED POLYMER MELTS}

\subsection{Boson Statistical Mechanics}

The partition function for a set of $N$ nonrelativistic bosons 
interacting with pair potential $V(r)$ reads
\begin{equation}
{\cal Z}=Tr'\{e^{-\beta\hat\ch_b}\ra,
\eqnum{2.1}
\label{eq:one}
\end{equation}
where $\beta=1/k_BT$ and the boson Hamiltonian operator is 
\begin{equation}
\hat\ch_b=\sum_{j=1}^N
{-\hbar^2\over 2m}
\nabla_j^2+{1\over 2}
\sum_{i\not=j}
V(|\r_i-\r_j|).
\eqnum{2.2}
\label{eq:two}
\end{equation}
The prime on the trace means that only 
symmetrized boson eigenfunctions are to be included in the 
partition sum, and we shall focus on particles in two 
space dimensions. This trace can be rewritten in
terms of a Feynman path integral by breaking up
$\exp[-\beta\hat\ch]$ into $M$ pieces $(M>>1)$,
\begin{equation}
e^{-\beta\hat\ch_b}=
e^{-\Delta\tau\hat\ch_b}
e^{-\Delta\tau\hat\ch_b}\cdots
e^{-\Delta\tau\hat\ch_b}\qquad 
(M\;\;\quad {\rm times})
\eqnum{2.3}
\label{eq:three}
\end{equation}
where $\Delta\tau M=\beta$. Upon inserting
complete sets of position states between 
various terms in the product 
and taking the limit $M\rightarrow\infty$, 
the boson partition function 
${\cal Z}_b$ may be expressed
as an integral over a set of polymer-like
trajectories $\{\r_j(\tau)=[x_j(\tau),y_j(\tau]\}$ 
in imaginary time
\cite{feynman}, 
\begin{eqnarray}
{\cal Z}_b&=&{1\over N!}
\sum_P
\prod_{j=1}^N
\cd\r_j(\tau)\exp
\left[
-{1\over 2}
{m\over\hbar}
\sum_{j=1}^N
\int_0^{\beta\hbar}
\left({d\r_j\over d\tau}
\right)^2\;d\tau\right. \nonumber \\
&&\left. -
{1\over\hbar}\sum_{i>j}
\int_0^{\beta\hbar}
V(|\r_i(z)-\r_j(z)|)d\tau\right]
\eqnum{2.4}
\label{eq:four}
\end{eqnarray}
The normalized  sum over permutations
$P$ insures that only boson eigenfunctions
contribute to the sum, where the trajectories obey
a type of periodic boundary condition,
\begin{equation}
\{\r_j(\beta\hbar\}=
P[\{\r_i(0)\}], 
\eqnum{2.5}
\label{eq:five}
\end{equation}
and the operator $P$ permutes the set of starting points
$\{\r_i(0)\}$.

An approximate picture of the polymer statistical
mechanics problem represented by (\ref{eq:four}) can be
constructed as follows \cite{feynman}: Provided interaction
effects are not too large, each ``polymer'' simply
diffuses in the imaginary time variable $\tau$,
\begin{equation}
\langle|\r_j(\tau)-\r_j(0)|^2\rangle
\approx 2{\hbar\over m}\tau
\eqnum{2.6}
\label{eq:wfive}
\end{equation}
where $\hbar/m$ plays the role of a diffusion
constant. When the temperature is high, only the
identity permutation contributes to Eq. (\ref{eq:four}).
When projected onto the $(x,y)$-plane,
the boson trajectories then behave like small {\it
ring} polymers of typical transverse size given by the
thermal deBroglie wavelength $\Lambda_T$,
\begin{eqnarray}
\Lambda_T^2&\equiv& 2\pi\hbar^2/mk_BT\nonumber \\ 
&\sim& \langle|\r_j(\tau)-\r_j(0)|^2
\rangle|_{\tau=\beta\hbar}.
\eqnum{2.7}
\label{eq:six}
\end{eqnarray}
At temperatures low enough so that $\Lambda_T
\got n^{-1/2}$, where $n$ is the particle number
density, complicated cyclic permutations appear,
as the trajectories coalesce to form much larger
rings. Feynman suggested in 1953 that the lambda
transition from a normal bulk liquid of He$^4$ to a superfluid is
associated with a dramatic proliferation in the
number and length of such cooperative ring
exchanges \cite{feynman91}. 

Now, following Pollock and Ceperley, consider what
happens when the excursions of the bosons in the 
$xy$-plane occur in a 
two-dimensional periodic box of size $D$ \cite{pollack,ceper}
(see Fig.~3).  The permutation requirement (\ref{eq:wfive})
describing periodic boundary conditions in the $\tau$-direction
remains in effect. At
high temperatures, virtually all ring polymers
return to their initial positions $\{\r_i(0)\}$ 
when $\tau=\beta \hbar$, and the spatial periodic
boundary conditions are unimportant. In the low
temperature limit, however, ring exchanges 
lead to huge composite trajectories which
typically wrap completely around the torus
embodied in the $(x,y)$-plane
periodic boundary conditions. The
collection of cyclic boson trajectories can be
classified by a dimensionless 
topological invariant, the vector 
winding number $\W$,
\begin{equation}
\W={1\over\sqrt{\Omega}}
\sum_{j=1}^N [\r_j(\beta\hbar)-\r_j(0)]
\eqnum{2.8}
\label{eq:seven}
\end{equation}
where $\Omega=D^2$ is the cross-sectional area of
a square box with periodic spatial boundary conditions. In
evaluating Eq. (\ref{eq:seven}), we imagine
that the $\{\r_j(\tau)\}$ pass smoothly into
neighboring periodic cells, without invoking the
periodic boundary conditions (see Fig.~3).
Consider the mean square winding number $\la
\W^2\ra$, where the angular brackets represent a
path integral weighted by the exponential factor
in Eq. (\ref{eq:four}) and divided by the boson partition
function. Ring polymers which do not wrap completely around
the torus make no contribution to Eq. (\ref{eq:seven}).
However, at temperatures low enough so that
$\Lambda_T>>n^{-1/2}$, virtually all trajectories
belong to a cycle with a nontrivial winding
number. We can then regard Eq. (\ref{eq:seven}) as a normalized 
$N$-step random walk with typical step size
$\Lambda_T$, and estimate that $\la\W^2\ra\approx
{N\Lambda_T^2\over\Omega}=2\pi 
n\hbar^2/mk_BT$, where $n=N/\Omega$ is the number 
density of bosons.
Pollock
and Ceperley showed that $\la \W^2\ra$ is in fact
{\it exactly} related to the {\it superfluid
density} $n_s$ of the equivalent boson system,
\begin{eqnarray}
\la\W^2\ra&=&
2n_s(T)\hbar^2/mk_BT\nonumber \\ 
&=&2n_s(T)(\hbar^2/m)\beta .
\eqnum{2.9}
\label{eq:eight}
\end{eqnarray}

The implications of this remarkable connection
between a {\it topological invariant} for a set of ring
polymers on a torus and the {\it superfluid
density} of the equivalent set of bosons is
summarized in Fig.~4, where $\la \W^2\ra$ is shown
as a function of $\beta$, which is proportional to
the number of ``monomers'' if the trajectories are 
viewed as  ring polymers.
Translational invariance of the boson system
in the absence of disorder
implies that $\lim_{T\rightarrow 0}n_s(T)=n$,
\cite{nozieres}, so we know that $\la \W^2\ra$
diverges linearly  with $\beta$ as
$\beta\rightarrow\infty$, with slope given exactly by
$2n\hbar^2/m$.
The existence of a sharp Kosterlitz-Thouless phase
transition in superfluid helium films, moreover, implies that
there must be a singularity in $\la \W^2\ra$: This
quantity {\it vanishes} below a critical value
$\beta_c$, and the universal jump discontinuity in
the superfluid density \cite{nelson77}, $\lim_{T\rightarrow
T_c^-}n_s(T)/T={2\over\pi} k_B{m\over\hbar^2}$,
implies the exact result \cite{pollack} 
\begin{equation}
\lim_{\beta\rightarrow\beta_c^+} \la
\W^2\ra={4\over\pi}.
\eqnum{2.10}
\label{eq:nine}
\end{equation}

Feynman  hoped that his ``cyclic ring
exchange'' picture of superfluidity could be used
to understand the lambda transition in He$^4$
which was a quite mysterious phenomenon in 1953
\cite{feynman,feynman91}. Here, the sophisticated understanding
of vortex unbinding transitions in He$^4$ films
developed in the past 25 years (using other methods)
has been used to
make a prediction about a topological quantity,
the winding number. 

Equation (\ref{eq:eight}) allows winding
number statistics extracted from computer
simulations of bosons with periodic boundary
conditions to be converted into measurements of
the superfluid density \cite{ceper}.
In can be used, in particular, to probe the reduction 
in the superfluid density of helium films due to 
substrate disorder near $T=0$. A disordered substrate 
potential maps onto randomness {\it correlated} along the 
imaginary time direction $\tau$ in the 2+1-dimensional 
world line picture of boson physics. The winding number 
fluctuations are reduced because this correlated 
randomness reduces the wandering of the boson world lines.

\subsection{Directed Polymer Melts}

One might think that the results described above
would be applicable, at least in principle, to
real ring polymers on the surface of a torus,
whose fluctuations include fusing together to form
large rings, similar to sulfer ring polymers in
equilibrium \cite{bellissent}. Unfortunately, the boson model
as applied to ring polymers is unrealistic: The
individual ``monomers'' within a ring, indexed by
the imaginary time coordinate $\tau$ in Eq. (\ref{eq:four}),
are noninteracting, so there is no intra-chain
self-avoidance. The only {\it inter}polymer
interactions, moreover, occur between monomers
with {\it same} imaginary time coordinate, which is 
also unrealistic for a melt of ring polymers.

The
physics of directed polymer melts, on the other
hand, is much closer to that of real bosons. Because
the polymers are directed, interpolymer interactions 
at the same ``imaginary time'' coordinate dominate the 
physics. Vortex
lines in superconductors provide an excellent
physical realization of this system, but examples
can also be found in polymer nematics in strong
external magnetic or electric fields \cite{nelsonphys}.
Consider a collection of $N$ vortex lines or
polymers in three space dimensions, labelled by
$x$, $y$ and $\tau$ (see Fig.~5). We assume that
these lines are stretched out on average along the
$\tau$ axis, so that they can be described by
single-valued trajectories $\{\r_j(\tau)\}$. The
partition function is a multidimensional path
integral, similar to Eq. (\ref{eq:four}), 
\begin{equation}
{\cal Z} = \prod_{j=1}^N
\int\cd\r_j(\tau)\exp[-F/T]
\eqnum{2.11{\rm a}}
\label{eq:tena}
\end{equation}
where
\begin{equation}
F={1\over 2} g\sum_{j=1}^N
\int_0^L
\left({d\r_j\over dz}\right)^2
d\tau +
\sum_{j=1}^N\int_0^L
U(\r_j)d\tau+
\sum_{i>j}\int_0^L
V(|\r_i-\r_j|)d\tau.
\eqnum{2.11{\rm b}}
\label{eq:tenb}
\end{equation}
Here, the mass $m$ in the boson problem has been
replaced by a line 
tension $g$, $\hbar$ has been replaced by the
temperature $T$
and the polymer system thickness $L$ plays the
role of $\beta\hbar$ for the bosons. Unless indicated
otherwise, we shall henceforth use units such 
that $k_B=1$. $V(r)$ is a
repulsive interparticle potential and $U(\r)$
represents the random  potential due to columnar pins 
(not shown in Fig.~5). Both
$V(r)$ and $U(\r)$ are independent of $\tau$. This
approximation is quite reasonable for directed
polymer melts and vortex lines in $2+1$
dimensions, provided $\la|{d\r\over dz}|^2\ra$, 
the mean square tilt away from the $z$ axis, is
small \cite{nelson93}.

The statistical mechanics of directed polymers differs
from that for bosons due to the absence of
periodic boundary conditions in the ``imaginary
time'' direction $\tau$: we assume free boundary
conditions in Eq. (\ref{eq:tena}), i.e., we integrate
freely over the starting and end points of the
polymers. Hence there is no sum over
permutations and no condition analogous to Eq.
(\ref{eq:five}). This change, however, is less severe than
one might think. Indeed, the directed polymer
partition function can be written in a form
similar to Eq. (\ref{eq:one}), \cite{nelson}
\begin{equation}
{\cal Z} =\prod_{j=1}^N
\int
d\r_j'\prod_{j=1}^N\int d\r_j \la\r'_1
\cdots\r'_N|e^{-\hat \ch L/T}|
\r_1\cdots\r_N\ra
\eqnum{2.12}
\label{eq:eleven}
\end{equation}
where
\begin{equation}
\hat \ch=-{T^2\over 2g}
\sum_{j=1}^N
\nabla_j^2+\sum_{j=1}^N
U(\r_j)+\sum_{i>j}
V(|\r_i-\r_j|),
\eqnum{2.13}
\label{eq:twelve}
\end{equation}
and the states $|\r_1\cdots\r_N\ra$ and
$\la\r'_1\cdots\r'_N|$ describe entry and exit
points for the polymers or vortices at the top and
bottom of the system. The Hamiltonian $\ch$ is 
identical to the Hamiltonian $\ch_b$ in Eq. (\ref{eq:two}),
provided we introduce a disorder potential 
$U(\r)$ to model the effect of, say, a random 
substrate on the bosons. Because the initial and
final states involve {\it symmetric} integrations
over the entry and exit points, only boson
eigenfunctions contribute to the statistical
mechanics defined by Eq. (\ref{eq:eleven}).  As,
$L\rightarrow\infty$, the physics will be
dominated by a bosonic ground state and bosonic
low-lying excitations, just as in Eq. (\ref{eq:one}).
Thus, the precise choice of boundary condition is irrelevant
in the ``thermodynamic limit'' of large system sizes.

Can one define meaningful ``winding number'' problems
for directed polymers with free ends? Because we
are no longer dealing with ring polymers, the
``winding number'' $\W$ in Eq. (\ref{eq:seven}) (with
$\beta\hbar$ replaced by $L$) no longer has a
precise topological meaning, even for periodic
boundary conditions in the space-like directions.
One can, however, still define and estimate the
order of magnitude of the fluctuations in
$\W$ as before. In
terms of quantities appropriate for directed
polymer melts, we have, up to constants of order unity
\begin{equation}
\la \W^2\ra\approx nTL/g,
\eqnum{2.14}
\label{eq:thirteenj}
\end{equation}
where $n$ is the areal number density of polymers
in a constant $\tau$ cross section. As
illustrated (in the absence of a disorder 
potential) in Fig.~4, we expect that this
quantity is nonzero for {\it all} values of
$L\equiv\beta\hbar>0$. Although the curves for
directed polymers and bosons should agree when $L$
and $\beta\hbar$ are large, there is no reason to
expect a sharp Kosterlitz-Thouless phase
transition for ``bosons'' with open boundary
conditions \cite{fisher}. The winding number variance
for real bosons  only vanishes
identically for $\beta <\beta_c$ because the
contribution to $\W$ from ``ring polymer''
trajectories which are not wrapped completely
around the torus is exactly zero.
There is no such condition for the open boundary conditions
appropriate to a directed polymer melt. 

\subsection{Winding Around an Obstacle}

A more interesting case is the winding number
distribution for many directed polymers in a
disorder potential and interacting with a
repulsive cylindrical obstacle. As discussed in
the Introduction, this problem has a direct
physical realization in the context of thermally
excited vortex lines. The combination of a
disorder potential and repulsive barrier is
illustrated in Fig.~1. We  model a repulsive
obstacle by modifying  the disorder potential in 
Eqs. (\ref{eq:tenb}) and (\ref{eq:twelve}), 
\begin{equation}
U(\r)\rightarrow U(\r)
+U_B[\r(\tau)],
\eqnum{2.15}
\label{eq:fourteen}
\end{equation}
where the barrier potential
$U_B(\r)$ (centered on the origin) is, e.g., 
\begin{equation}
U_B(r)=\left\{\matrix
{\infty,&\quad r<a\cr
0,&\quad {\rm otherwise}.}\right.
\eqnum{2.16} \label{eq:fifteen} 
\end{equation}
This problem is a natural generalization to many directed 
lines of the original multiply connected geometry
problem for single polymers 
of Edwards \cite{edwards}  and of Prager and Frisch
\cite{prager}. The (scalar) winding number with respect to the
obstacle at the origin is now 
\begin{equation}
\W=\sum_{j=1}^N
\int_0^L
d\tau \A[\r_j(\tau)]
\cdot {d\r_j\over d\tau},
\eqnum{2.17}
\label{eq:sixteen}
\end{equation}
where the ``vector potential'' $\A(r)$ is \cite{edwards}
\begin{equation}
\A(\r)=
{\hat{\Btau}\times\r\over 2\pi r^2}.
\eqnum{2.18}
\label{eq:seventeen}
\end{equation}
The boundary conditions along the $\Btau$
direction can be periodic or open \cite{gross}. Periodic
boundary conditions would imply integer winding
numbers, but the noninteger winding numbers which
characterize open boundary conditions are of
direct physical interest for vortices, as we now explain. 

A current in the $\Btau$ direction through a
cylindrical obstacle couples directly to the
winding number of vortices outside in Type II
superconductors. To see this, assume that the
current is entirely confined to the cylinder,
i.e., it cannot penetrate into the remainder 
of the sample.  This ``confined current
approximation'' should be applicable to a tube
of columnar pins filled with pinned vortices
in the bose glass phase, provided its
conductivity is much higher than that of the flux
liquid outside. Essentially infinite conductivity
can be arranged in the tube because irradiation to produce
columnar defects with matching field $\sim$~4 Tesla
typically shifts the irreversibility temperature by 
5--20$^\circ$K. 
Choosing a temperature {\it above} the irreversibility
line of the lightly irradiated region outside the tube 
but {\it below} the Bose glass transition inside insures
that essentially all the current is confined to the tube. 
Under these conditions the
additional free energy $F$ of the vortices due
the current is given by \cite{gennes}
\begin{equation}
\delta F=-{1\over 4\pi}
\int^{\prime}
d^2r\int_0^Ld\tau
\H_{\rm ext} (\r)\cdot
\B_\perp(\r,\tau)
\eqnum{2.19}
\label{eq:eighteen}
\end{equation}
where $\H_{\rm ext}(\r)$ is the field due to the
extra current $I$ through the wire
\begin{equation}
\H_{\rm ext}(\r)=
{2 I\over c}{\hat {\Btau}\times\r\over
r^2},
\eqnum{2.20}
\label{eq:nineteen}
\end{equation}
and the prime on the integration means that the
region inside the obstacle is excluded. The 
perpendicular component of the magnetic field 
$\B_\perp$ due to the vortices obeys the
anisotropic London equation \cite{kogan}
\begin{equation}
\B_\perp(\r,\tau)=
\phi_0\sum_{j=1}^N
{d\r_j\over d\tau}
\delta^{(2)}[\r-\r_j(\tau)]+
{M_\tau\over M_\perp}\lambda^2\nabla_\perp^2
\B_\perp(\r,\tau)+\lambda^2
\partial_\tau^2\B_\perp(\r,\tau),
\eqnum{2.21}
\label{eq:twenty}
\end{equation}
where $\phi_0$ is the flux quantum, ${M_\tau\over
M_\perp}$ is the mass anisotropy and $\lambda$ is
the screening length in the direction
perpendicular to $\hat{\Btau}$. Upon substituting
Eqs. (\ref{eq:nineteen}) and (\ref{eq:twenty}) into 
(\ref{eq:eighteen}), integrating by 
parts and noting that 
$\left({M_\tau\over M_\perp} \lambda^2
\nabla_\perp^2+\lambda^2\partial_\tau^2
\right) \H_{\rm ext}(\r)=0$ in the domain of 
integration, we find that 
\begin{eqnarray}
\delta F&=&-{I\phi_0\over  c}
\sum_{j=1}^N
\int_0^L d\tau
{\hat{\Btau}\times\r_j\over 2\pi r_j^2}
\cdot{d\r_j\over d\tau} \nonumber \\ 
&\equiv& -\gamma W,
\eqnum{2.22}
\label{eq:twentyone}
\end{eqnarray}
where we have defined a winding number coupling
constant $\gamma$,
\begin{equation}
\gamma=I\phi_0/ c.
\eqnum{2.23}
\label{eq:twentytwo}
\end{equation}

Consider now the mean winding number $\la W\ra$
induced by the current for vortices outside the
obstacle. Upon incorporating an additional factor 
$e^{-\delta F/T}$ into the Boltzmann weight in  (\ref{eq:tena}),
we have, upon expanding $\langle W\rangle$
to leading order in the current, 
\begin{equation}
\la W\ra={\phi_0\over 2\pi cT}
\la W^2\ra|_{I=0}
I.
\eqnum{2.24}
\label{eq:twentythree}
\end{equation}
Thus the equilibrium winding number {\it fluctuations} 
$\la W^2\ra|_{\RI=0}$ determine the linear response
of the net winding number to an external current.

\section{WINDING NUMBER AND QUANTUM MECHANICS: SINGLE POLYMER IN A CYLINDRICAL SHELL}

We now study the winding number fluctuations of a single directed
polymer,  moving on
average along the $\tau$ axis, and interacting
with a repulsive cylindrical obstacle of radius
$a$ centered on the origin of an $(x,y)$ plane
perpendicular to $\hat{\Btau}$. As discussed above,
many of the results are directly applicable to
vortex filaments in superconductors. We use the
original method of Edwards \cite{edwards}, which maps the
generating function for winding numbers onto the
quantum mechanics of particles interacting with a
solenoidal vector potential. Although winding of a
single polymer is simple and relatively well
understood \cite{edwards,prager,gross,rudnick,drossel}, we review the basic results
to illustrate the method and discuss effects due to 
columnar pins 
outside the obstacle. We use the
{\it directed} polymer notation discussed at the
end of section~II because the neglect of
self-avoidance is justified in this case. The
single line results would, however, apply in
principle to a ``phantom'' polymer without
self-avoidance in two dimensions interacting with
a repulsive disk and a random substrate potential.
We defer  new results
for {\it many} directed polymers to section IV. We shall, for
simplicity, usually impose periodic boundary
conditions in the $\tau$ direction. This choice 
should not affect the results
for $\la W^2\ra$ in the limit of large sample
thicknesses $L$. 
For a systematic study of the effects of
different boundary conditions in the context of the boson
mapping, see ref. \cite{tauber}.

\subsection{Winding Number Formalism}

Consider the winding number (\ref{eq:sixteen}), specialized to
the case of a single directed polymer with
trajectory $\r(\tau)$,
\begin{equation}
W=\int_0^L
d\tau\A
[\r(\tau)]\cdot \dot {\r}(\tau)
\eqnum{3.1}
\label{eq:tone}
\end{equation}
with $\A(\r)={\hat{\Btau}\times\r\over 2\pi r^2}$
and $\dot{\r}(\tau)=d\r/d\tau$. Note that
$\A(\r)$ is the electro-magnetic vector
potential that would have been present had we
replaced the cylindrical obstacle by a solenoid 
enclosing one unit of magnetic flux. If we then regard
$\tau$ as time, our expression for $W$ becomes the
quantum mechanical Aharonov-Bohm phase accumulated
by a particle of unit charge traversing a path
$\r(\tau)$, due to its interaction with the
solenoid. As noted by Aharonov and Bohm \cite{aharonov}, this
phase is indeed the winding number, which is an
integer for periodic boundary conditions, but
can assume arbitrary values otherwise. 

Our interest is, of course, not in the winding
number corresponding to a single trajectory, but
rather in the statistics of winding numbers. More
concretely, we wish to study the distribution
of $W$, averaged over all paths, with the
appropriate Boltzmann weight.
The winding number variance for a single polymer with line tension $g$ is

\begin{equation}
\la W^2\ra=
{
\int\cd\r(\tau) \left[
\int d\tau \left[
\dot\r\cdot\A(\r(\tau) \right]^2
e^{-\int_0^L
{\cal L}_0
[\r(\tau),\dot\r(\tau)]
d\tau/T}\right] \over
\int
\cd\r(\tau)e^{
-\int_0^L{\cal L}_0
[\r(\tau),\dot\r(\tau)]
d\tau/T}},
\eqnum{3.2}
\label{eq:ttwo}
\end{equation}
where the free energy density
\begin{equation}
{\cal L}_0={1\over 2}
g\left({d\r\over d\tau}\right)^2
+U(\r)
\eqnum{3.3}
\label{eq:tthree}
\end{equation}
plays the role of a Lagrangian in the imaginary
time path integral formulation of quantum
mechanics. Here, $U(\r)$ includes the potential
due to the columnar (i.e. $\tau$-independent)
disorder and a part representing the 
excluded volume interaction with the cylinder, 
as in Eq. (\ref{eq:fourteen}).

Upon defining a new function
\begin{equation}
{\cal L}_\nu
[\r(\tau),\dot\r(\tau)]={\cal L}_0
-iT\nu{d\r\over d\tau}\cdot \A[\r(\tau)],
\eqnum{3.4}
\label{eq:tfour}
\end{equation}
we have
\begin{equation}
\la W^2\ra=-
{\partial^2\over\partial\nu^2}
\ln
{\cal Z}(\nu)|_{\nu=0},
\label{eq:tfive}
\eqnum{3.5}
\end{equation}
where
\begin{equation}
{\cal Z}(\nu)=\int\cd
\r(\tau)\exp\left[
-{1\over T}\int_0^L
{\cal L}_\nu[\r(\tau),\dot{\r}(\tau)]d\tau\right].
\eqnum{3.6}
\label{eq:tsix}
\end{equation}

It is straightforward using standard path integral
manipulations to express ${\cal Z}(\nu)$ in terms
of $\hat \ch(\nu)$, the Hamiltonian operator associated
with $\cl_\nu$,
\begin{equation}
\hat\ch(\nu)=
{T^2\over 2g}
\left[ {1\over i}
\Bnabla+\nu\A(\r) \right]^2
+U(\r),
\eqnum{3.7}
\label{eq:tseven}
\end{equation}
specifically,
\begin{equation}
{\cal Z}(\nu)=
{\rm Tr}
\left\{ e^{-\hat\ch(\nu)L/T}
\right\},
\eqnum{3.8}
\label{eq:teight}
\end{equation}
where the trace means we have now imposed periodic
boundary conditions. For a system with cross-sectional 
area $\Omega$ the mean square winding
number itself is
\begin{equation}
\la W^2\ra={L\over \Omega}
{\partial\over\partial\nu}
\int d\r\la j_\nu(\r)\ra
\cdot\A(\r)|_{\nu=0},
\eqnum{3.9{\rm a}}
\label{eq:tninea}
\end{equation}
\begin{equation}
\la W^2\ra=
{L\over T}\left\langle {
\partial^2\hat\ch(\nu)\over\partial^2\nu}\right\rangle_{\nu=0}
-{1\over T^2}\int_0^L d\tau\int_0^L
d\tau'\left\langle
\hat T\left[{\partial \hat\ch\over\partial\nu}
(\tau){\partial \hat\ch\over\partial \nu}
(\tau')\right]\right\rangle_{\nu=0},
\eqnum{3.9{\rm b}}
\label{eq:tnineb}
\end{equation}
where
\begin{equation}
\j_\nu(\r)\equiv -{1\over T}{\partial\hat\ch\over 
\partial \A}={T\over g}
\left[{1\over i}{\bf \nabla}+\nu
\A(\r)\right]
\eqnum{3.10}
\label{eq:tten}
\end{equation}
is the current operator, $\la \hat O\ra\equiv
{1\over Z}
\tr \;\{\hat O e^{-\beta \hat \ch}\}$, 
$\hat T$ is the time ordering operator, and time
dependence of operators is defined according to
the Heisenberg picture in imaginary time $\hat
O(\tau) \equiv e^{-\hat\ch\tau}\hat O e^{\hat \ch\tau}$.

Equations (\ref{eq:tninea},\ref{eq:tnineb}) are central to our discussion. On
the left-hand side they both have the fluctuations
in the winding number. On the right-hand side, the
first equation has the derivative of the
persistent, thermodynamic, current due to a
fictitious quantum particle flowing in a multiply
connected geometry threaded by an Aharonov-Bohm
flux. The derivative is taken with respect to the
particle ``charge'' $\nu$, and the current is
weighted by the vector potential. Equation
(\ref{eq:tnineb}) expresses this derivative in terms of
one- and two-time particle correlation functions.
An analogous equation for {\it many} directed
polymers in terms of boson correlation functions
will be given in section IV.
Information known from  studies of quantum mechanical 
particles and many particle boson systems regarding 
these
correlation functions will allow us to determine 
$\la W^2\ra$.   

Equations (\ref{eq:tninea},\ref{eq:tnineb}) are expressions for the variance
of the winding number $W$. We now briefly discuss the
probability that the winding number $W$ equals
$n$, denoted by $P_W(n)$. This probability is,
\begin{equation}
P_W(n)=
{\cp_W(n)\over\sum_n\cp
_W(n)},
\eqnum{3.11}
\label{eq:televen}
\end{equation}
where
\begin{eqnarray}
\cp_W(n)&=&
\int D\r(z)\delta
\left[\int d\tau \A[\r(z)]\cdot \dot{\r}-n\right]
e^{-\int_0^L \cl_0[\r(z),\dot{\r}(z)]dz/T}
\nonumber \\ 
&=& {1\over 2\pi}
\int d\nu\int D\r(\tau)
e^{-{1\over T}\int_0^L \cl[\r(\tau),\dot{\r}(\tau)]d\tau+i\nu
\int_0^L d\tau\A[\r(\tau)]\cdot\dot{\r}-i\nu n}
\nonumber \\ 
&=&{1\over 2\pi} \int d\nu
{\cal Z}(\nu)e^{-i\nu n}
\eqnum{3.12}
\label{eq:ttwelve}
\end{eqnarray}
with ${\cal Z}(\nu)$ given by (\ref{eq:tsix}). Equation
(\ref{eq:televen}) is again a mapping of the statistical
property we are interested in onto a quantum
mechanical problem, albeit a less transparent one:
the probability $P_W(n)$ is proportional to the
Fourier transform, with respect to $\nu$, of the
quantum mechanical partition function of a
particle of charge $\nu$ on an Aharonov-Bohm ring
threaded by a flux $1/2\pi$. For a related discussion, including 
an earlier derivation of Eq. 
(\ref{eq:ttwelve}), see Ref. \cite{comtet}.

The normalization of the probability distribution
function $\cp_W(n)$ is a matter of some subtlety.
In the second line of Eq. (\ref{eq:ttwelve}) we used the
identity,
\begin{equation}
\delta\left[\int d\tau
\A[\r(\tau)]\cdot\dot{\r}-n\right]=
{1\over 2\pi}\int d\nu e^{i\nu
\left(\int_0^L d\tau \A[\r(\tau)]\cdot\dot{\r}-n\right)}.
\eqnum{3.13}
\label{eq:tthirteen}
\end{equation}
The limits of the $\nu$ integration are determined
by the allowed values of the winding number $\int
d\tau\A[\r(\tau)]\cdot\dot\r$. Since we 
assume periodic boundary conditions, the allowed trajectories
are closed $(\r(0)=\r(L))$, the winding number is
an integer, and the integral over $\nu$ can be
taken between 0 and $2\pi$. However, if the
allowed trajectories are not closed, the winding
number can be non-integer, and the integral over
$\nu$ should be taken between $-\infty$ and
$+\infty$. For periodic boundary conditions, the integration
domain $\nu\in[0,2\pi]$ leads to 
$\sum_n\cp_W(n)=1$. Although we have focused on periodic
boundary conditions, 
in the limit $L\rightarrow\infty$ one
expects the difference between periodic and open
trajectories to vanish. 

\subsection{A Single Polymer on a Thin Cylinder}

We now specialize to  a  polymer, whose
motion in the $x-y$ plane is confined to a ring of
radius $a$, but is stretched out  along the
$\tau$ axis. The polymer's Lagrangian is given by
Eq. (\ref{eq:tthree}) and we add to  $U(\r)$ a piece which
confines the particle to an {\it annulus} just outside
the cylindrical barrier of radius $a$. 
Our previous discussion maps its
random walk around the ring onto the thermodynamics
of a quantum charged particle on an Aharonov-Bohm
ring. The solenoid's vector potential is then 
$\A={\hat{\Btheta}\over 2\pi a}= $~constant,
where $\hat{\Btheta}$ is a unit vector around the ring.

In the absence of disorder, we easily reproduce 
the expected results. According to Eq. (\ref{eq:tnineb}) 
the winding number variance is
\begin{equation}
\la W^2\ra=
{LT\over g(2\pi a)^2}
[1-2 L\la E(\nu=0)\ra]
\eqnum{3.14}
\label{eq:tfourteen}
\end{equation}
where $\la E(\nu=0)\ra$ is the average kinetic
energy of a chargeless particle on a ring. For a
very large $L$ the average energy is, to
exponential accuracy, the energy of the ground
state, i.e., 0, and $\la W^2\ra={LT\over (2\pi
a)^2g}$. As expected for a random walk, the
variance of the winding number is proportional to
the number of steps taken (i.e., proportional to
$L$). The probability distribution $P_W(n)$ is a
Gaussian in this limit, as expected \cite{drossel}. When
$L\rightarrow 0$, 
$\la E\ra\rightarrow {1\over 2L}$, and $\la
W^2\ra\rightarrow 0$.

The presence of a disorder potential $U(\r)$
modifies these results in an interesting way. The
effect of disorder on
\begin{eqnarray}
\la j_\nu\ra&=&
{1\over T}
\left\langle{\partial\hat\ch\over
\partial A}\right\rangle\nonumber \\
&=&
{-2\pi a\over L}{\partial\over\partial \nu}
\ln {\cal Z}(\nu),
\eqnum{3.15}
\label{eq:tfifteen}
\end{eqnarray}
i.e., the persistent current in a mesoscopic ring
threaded by an Aharonov-Bohm flux, is well known:
the disorder introduces a length scale, $\xi$, the
quantum mechanical localization length of the
ground state wave function. As
$L\rightarrow\infty$,
\begin{equation}
Z(\nu)\approx\exp[-L\epsilon_0(\nu)/T],
\eqnum{3.16}
\label{eq:tsixteen}
\end{equation}
where $\epsilon_0(\nu)$ is the ground state
energy. As a crude, but illuminating 
model of localization, consider
a {\it single} narrow  trap of depth $U_0$ running up the
side of the cylinder. The periodic boundary
conditions around the cylinder then lead to a
tight-binding model type result for the ground
state energy as a function of $\nu$,
\begin{equation}
\epsilon_0(\nu)=-U_0-2t_1\cos(\nu)-2t_2\cos(2\nu)
-\cdots 
\eqnum{3.17}
\label{eq:tseventeen}
\end{equation}
where the couplings $t_n$ represent the matrix
elements for tunneling around the cylinder $n$
times, $t_n\sim e^{-2\pi a n /\xi}$. 
For a square well of depth $U_0$ and size $b$ with 
$U_0>>T^2/2gb^2$ we have (see, e.g. [11])
$\xi\approx T/\sqrt{2gU_0}$ and $t_1\approx
{T^2\over gb^2}e^{-2\pi a/\xi}$. 
Upon using only
the first two terms to evaluate (\ref{eq:tfifteen}), we find
\begin{equation}
\la j(\nu)\ra=
{4\pi at_1\over T}\sin\nu,
\eqnum{3.18}
\label{eq:teighteen}
\end{equation}
and it follows from Eq. (\ref{eq:tninea}) that the mean square
winding number behaves as
\begin{eqnarray}
\la W^2\ra&=&
2t_1L/T\nonumber \\ 
&\sim& e^{-2\pi a/\xi}L.
\eqnum{3.19}
\label{eq:tnineteen}
\end{eqnarray}
Although the variance is still proportional to
$L$, the coefficient $2t_1/T$ 
which multiplies $L$ vanishes
{\it exponentially} fast as $a\rightarrow\infty$.

\subsection{A Single Polymer on a Thick Cylinder}

In the previous section we considered the
statistics of winding numbers associated with the
motion of a polymer on a thin cylinder or, equivalently, a
random walker on a ring. For a more general 
cylinder of outer radius $R$, there are two
extreme cases, depending on the
ratio of the cylinder width $R-a$ and the typical
transverse distance the polymer crosses in ``time'' $L$,
namely $\sqrt{LT/g}$. In the previous section this
ratio was zero. Suppose now that  
this ratio to be small, but nonzero, i.e.
$0<{g(R-a)^2\over TL}<<1$.
When using eigenfunctions of $\hat\ch$ to evaluate 
winding
numbers, this condition implies that the partition
function has significant contributions only from
one radial mode, the one contributing the lowest
energy. In this limit, the density profile in the
radial direction can be approximated by a
constant, and an analysis similar to that
following Eq. (\ref{eq:tfourteen}) leads, for a 
disorder-free sample in the limit
$L\rightarrow\infty$, to
\begin{equation}
\la W^2\ra=
{LT\over g\pi(R^2-a^2)}
\log {R\over a}.
\eqnum{3.20}
\label{eq:ttwenty}
\end{equation}

The opposite limit, in which the cylinder width is
essentially infinite, was considered, also in the
absence of disorder, in Refs. \cite{rudnick} and 
\cite{drossel}. Here, we
confine ourselves to a simplified discussion of
$\la W^2\ra$ using the quantum formalism.
According to Eq. (\ref{eq:ttwo}), the variance $\la W^2\ra$
can be written,
\begin{equation}
\la W^2\ra=
\int_0^L d\tau\int_0^L d\tau'
\left\langle{dr_i(\tau)\over d\tau}
{dr_j(\tau')\over d\tau'}
A_i[\r(\tau)]A_j[\r(\tau')]
\right\rangle.
\eqnum{3.21}
\label{eq:ttwentyone}
\end{equation}
Upon approximating the average in the integrand,
\begin{eqnarray}
\left\langle
{dr_i(\tau)\over d\tau}
{dr_j(\tau')\over d\tau'}
A_i[\r(\tau)]A_j
[\r(\tau)]\right\rangle&\approx&
\left\langle{
dr_i(\tau)\over d\tau}
{dr_j(\tau')\over d\tau'}\right\rangle
\la A_i[\r(\tau)]A_j[\r(\tau')]\ra\nonumber \\ 
&\approx&
{2T\over g} \delta(\tau-\tau')
\la A^2[\r(\tau)]\ra,
\eqnum{3.22}
\label{eq:ttwentytwo}
\end{eqnarray}
we have
\begin{equation}
\la W^2\ra=
{2T\over g}
\int_0^L d\tau\int
d^2 r\cp(\r,\tau)A^2(\r),
\eqnum{3.23}
\label{eq:ttwentythree}
\end{equation}
where $\cp(\r,\tau)$ is the probability of finding
the random walker at position $\r$ at height
$\tau$. For an infinitely thick cylinder, we have
$P(\r,\tau)={g\over 2\pi T\tau}e^{-gr^2/2T\tau}$
and $\la W^2\ra\propto (\log L)^2$. The thermal
averages in Eq. (\ref{eq:tninea},\ref{eq:tnineb}) now include many
eigenstates.

The effect of disorder can also be understood by a
similar analysis. Again, disorder introduces a
localization length $\xi$, characterizing the
eigenstates of the Hamiltonian $H(\nu)$ As long as
$\xi$ is larger than the cylinder's size, the
effect of the disorder is weak. When $\xi$ is
smaller than the cylinder's outer radius, but
larger than $2\pi a$, its inner circumference, the effective
outer radius of the cylinder becomes $\xi$.
Points on the plane whose distance from the obstacle
is larger than $\xi$ support mostly eigenstates
that are insensitive to $\nu$. Thus, they do not
contribute to the winding number. Particles that
start their way at such points never make it to
the hole. Therefore, the winding number for such a
system becomes,
\begin{equation}
\la W^2\ra\propto
\left\{ \matrix{
(\log L)^2&\quad{\rm for}&\quad {TL\over
g}<\xi^2,&\quad \xi> 2\pi a,\cr\cr
{L\over\xi^2}\log\left({\xi\over a}\right)
&\quad{\rm for}&\quad {TL\over
g}>\xi^2,& \quad \xi >2\pi a}\right. . 
\eqnum{3.24}
\label{eq:ttwentyfour}
\end{equation}
Finally, when $\xi<2\pi a$ the contribution to the
winding number comes from a 1-D strip around the
hole, and the 1-D result applies, namely $\la
W^2\ra\propto  e^{-2\pi a/\xi}L$. To summarize,
for an infinitely wide cylinder in the long-time limit
disorder makes the winding number increase, for
particles that start their random walk 
close enough to the hole. In the absence of
disorder the particles diffuse away from the hole,
and thus end up with a small rate of winding.
Disorder prevents them from diffusing away from
the hole, and increases the winding number.

\section{Winding Statistics of Many
Polymers}

We now turn to discuss a system with many polymers
(or flux lines). As discussed in section 2, the
two-dimensional Hamiltonian which describes the physics  
is that of interacting bosons, interacting
both mutually and with randomly placed pinning
centers. The pinning centers we discuss here are
assumed to result from columnar defects, and thus
create a $\tau$-independent potential.

The Hamiltonian  is a
generalization of Eq. (\ref{eq:twelve}), namely,
\begin{equation}
\hat\ch={T^2\over 2g}
\sum_{j=1}^N
\left[
{1\over i}\Bnabla_j+\nu\A(\r_j)\right]^2
+\sum_{j=1}^N U(\r_j)+
\sum_{i>j} V(|\r_i-\r_j|)
\eqnum{4.1}
\label{eq:fone}
\end{equation}
where $V$ is the interaction potential and $U$
includes both disorder and the interaction with a
repulsive cylinder centered on the origin. Note
that neither $V$ nor $U$ depend on $\nu$, which
determines only the coupling of the bosons to the
Aharonov-Bohm flux. The phase diagram of the
Hamiltonian (\ref{eq:fone}) includes several phases, most
notably a superfluid phase and an insulating bose glass
phase in the limit $L\rightarrow\infty$. 
The properties of these phases are
reflected in the winding number distributions. The
winding number for many directed lines is given by 
Eq. (\ref{eq:sixteen}),
and the many particle generalization of Eq. (\ref{eq:tnineb}) is
\begin{eqnarray}
\la W^2\ra={TL\over g}
\int d^2r\la n(\r)\ra A^2(r)&-&\int_0^L
d\tau\int_0^L d\tau'\int
d^2r\int d^2r'\la \hat T[j_\alpha(\r,\tau)
j_\beta(\r',\tau')]\ra_{\nu=0}\nonumber \\ 
&&\quad\qquad\qquad\times A_\alpha(\r)A_\beta(\r').
\eqnum{4.2}
\label{eq:ftwo}
\end{eqnarray}

Several comments are in place regarding Eq. (\ref{eq:ftwo}).
First, the density operator
$n(\r)=\sum_i\delta(\r-\r_i)$ describes the {\it
total} density of particles (rather than, e.g.,
the superfluid density). Second, the
current-current correlation function should be
calculated for $\nu=0$, i.e., the current operator
is $j_\alpha(\r)=
(1/2g)\sum_i[\delta(\r-\r_i)p_\alpha+p_\alpha\delta(\r-\r_j)]$.
Finally, the interpretation of the second term becomes
clearer upon Fourier transformation. After Fourier analysis,
the imaginary-time current-current correlation
function appearing on the right-hand side of Eq.
(\ref{eq:ftwo}) becomes the linear response function
$\chi_{\alpha\beta}(\q,\omega=0)$ defined by
$j_\alpha(\q,\omega=0)=\sum_\beta
\chi_{\alpha\beta}(\q,\omega=0)$
$A_\beta(\q,\omega=0)$, i.e., the function
describing the current response to a weak,
time-independent vector potential \cite{mahan}.

The current-current correlation function 
$\chi_{\alpha\beta}(\r,\tau)=\la \hat T[
j_\alpha(\r,\tau)j_\beta({\bf 0},0]\ra$ in Eq. (\ref{eq:ftwo})
may be decomposed in longitudinal and transverse
parts. The corresponding Fourier decomposition
reads
\begin{equation}
\chi_{\alpha\beta}(\r,\tau)=
{1\over\Omega L}
\sum_{\q,\omega}
e^{i\q\cdot\r-i\omega\tau}
\left[
\chi_\parallel (\q,\omega)
{q_\alpha q_\beta \over q^2}+
\chi_\perp (\q,\omega)
\left(\delta_{\alpha\beta}-
{q_\alpha q_\beta\over q^2}\right)\right].
\eqnum{4.3}
\label{eq:fthree}
\end{equation}
However, the vector potential
$\A(\r)={\hat{\Btau}\times\r\over
2\pi r^2}$ for winding around an 
obstacle is purely transverse, so only
$\chi_\perp(\q,\omega)$ contributes to the winding
number. Upon passing to the limit of large system
dimensions in the $xy$-plane, and rewriting the
second term of (\ref{eq:ftwo}) in Fourier space, we have
\begin{equation}
\la W^2\ra=
{TL\over g}\int d^2r
\la n(\r)\ra A^2(\r)-
L\int {d^2q\over (2\pi)^2}
\chi_\perp(\q,\omega=0)
|\hat{\A}(\q)|^2
\eqnum{4.4}
\label{eq:ffour}
\end{equation}
where $\hat\A(\q)$ is the Fourier transform of
$\A(\r)$.
\begin{equation}
\hat\A(\q)=
{-i\hat\tau\times\q\over |q|^2}.
\eqnum{4.5}
\label{eq:ffive}
\end{equation}
Suppose the directed polymers are confined in an
annulus of outer radius $R>>a$, where $a$ is the
radius of the cylindrical obstacle. As
$R\rightarrow\infty$, we can neglect any
distortion of the line density due to the obstacle
and approximate $\la n(r)\ra$ by its average value, $n$, far
from the inner boundary. The first term of Eq.
(\ref{eq:ffour}) then behaves like 
${TL\over 2\pi g} n\ln(R/a)$, i.e., it diverges
logarithmically with coefficient proportional to
the average density of lines. Because
$|\hat\A(\q)|^2=1/q^2$, the integral in the second
term is dominated by the behavior of
$\chi_\perp(q,\omega=0)$ in the limit
$q\rightarrow 0$. Upon imposing upper and lower
Fourier cutoffs of $a^{-1}$ and $R^{-1}$, we {\it again}
find a logarithmically diverging contribution 
to the winding number, and the mean 
square winding number in the large $R$ limit is 
\begin{equation}
\la W^2\ra=
{TL\over 2\pi g} n\ln(R/a)-
{L\over 2\pi}
\left[
\lim_{q\rightarrow 0}\chi_\perp(q,\omega=0)
\right]\ln (R/a).
\eqnum{4.6}
\label{eq:fsix}
\end{equation}
However, the transverse response function
$\chi_\perp(q,\omega)$ is well known to be related
to the {\it normal} density $n_n$  of
the equivalent system of interacting bosons \cite{nozieres,tauber}
\begin{equation}
\lim_{\q\rightarrow 0}
\chi_\perp(\q,\omega=0)=
{T \over g} n_n.
\eqnum{4.7}
\label{eq:fseven}
\end{equation}
Upon defining the superfluid number density
$n_s=n-n_n$, we are led to our final result,
namely
\begin{equation}
\la W^2\ra=
{TL\over 2\pi g} n_s\ln(R/a).
\eqnum{4.8}
\label{eq:feight}
\end{equation}
Equation (\ref{eq:feight}) is an exact relation, valid in the
limit $R>>a$, between winding number fluctuations
and the superfluid density of the equivalent boson
system. When reexpressed in terms of the boson 
parameters of section~2 it reads
\begin{equation}
\la W^2\ra=
{n_s\hbar^2\over 2\pi mk_BT}
\ln(R/a),
\eqnum{4.9}
\label{eq:fnine}
\end{equation}
which  is similar in some respects to the
Pollock-Ceperley result (\ref{eq:eight}) for boson world
lines winding around a torus \cite{equation}. Unlike the
Pollock-Ceperley formula, however, Eq. (\ref{eq:feight}) applies
directly to a real physical system, namely
``directed polymer melts'' composed of vortex
lines in Type II superconductors with columnar
defects. As discussed in section 2, 
$\la W^2\ra$ determines the linear response of the 
net winding number to a current through the obstacle.

According to Eq. (\ref{eq:feight})
the winding number fluctuation around an
obstacle for many interacting directed polymers
is predicted to diverge linearly with the length
of the obstacle and logarithmically with the
cross-sectional area, as one might guess by
multiplying the result (\ref{eq:ttwenty}) for a single polymer in a 
moderately thick cylinder by
$N$. The meaning of the coefficient becomes
clearer if we first define a ``thermal deBroglie
wavelength`` $\Lambda_T$ for the polymers by
\begin{equation}
\Lambda_T^2={2\pi TL\over g},
\eqnum{4.10}
\label{eq:ften}
\end{equation}
i.e., the directed polymer analogue of Eq. (\ref{eq:six})
with the usual identifications $\hbar\rightarrow
T$, $\hbar/T\rightarrow L$ and $m\rightarrow g$.
This length is a measure
of the transverse wandering distance of the
polymers as they traverse the sample. Equation
(\ref{eq:feight}) then becomes
\begin{equation}
\la W^2\ra=\left( {\Lambda_T\over 2\pi}\right)^2
n_s\ln(R/a),
\eqnum{4.11}
\label{eq:feleven}
\end{equation}
showing that only a part of the ``superfluid
fraction'' $n_s$ of lines, i.e., those within a
transverse wandering distance of the obstacle,
contribute to the winding number fluctuations.
Correlated disorder along the $\hat \tau$ axis
should decrease the winding number fluctuations.
Equation (\ref{eq:feleven}) shows that this reduction is given
entirely by the corresponding reduction in the
superfluid density. The reduction in $n_s$  for
directed lines subjected to  various kinds of correlated disorder 
is calculated explicitly in
Ref. \cite{tauber}.

\acknowledgements

It is a pleasure to acknowledge helpful conversations 
with D. Bishop and D. Ceperley. 
Work by DRN was supported by the National Science Foundation, 
primarily by the MRSEC program through Grant DMR--9400396
and in part through Grant DMR--9417047.


*To appear in Proceedings of the XIV Sitges 
Conference, ``Complex Behavior of Glassy Systems,''
June 10-14, 1996, edited by M. Rubi.

\newpage
\parindent=15pt
\centerline{FIGURES}
FIG. 1. \ Slab of Type II superconductor with a dense array of 
columnar pins confined to a cylinder with a dilute concentration
of columnar defects outside. Vortices inside the cylinder are 
in the Bose glass phase, while those outside are free to move 
and entangle in a flux liquid.

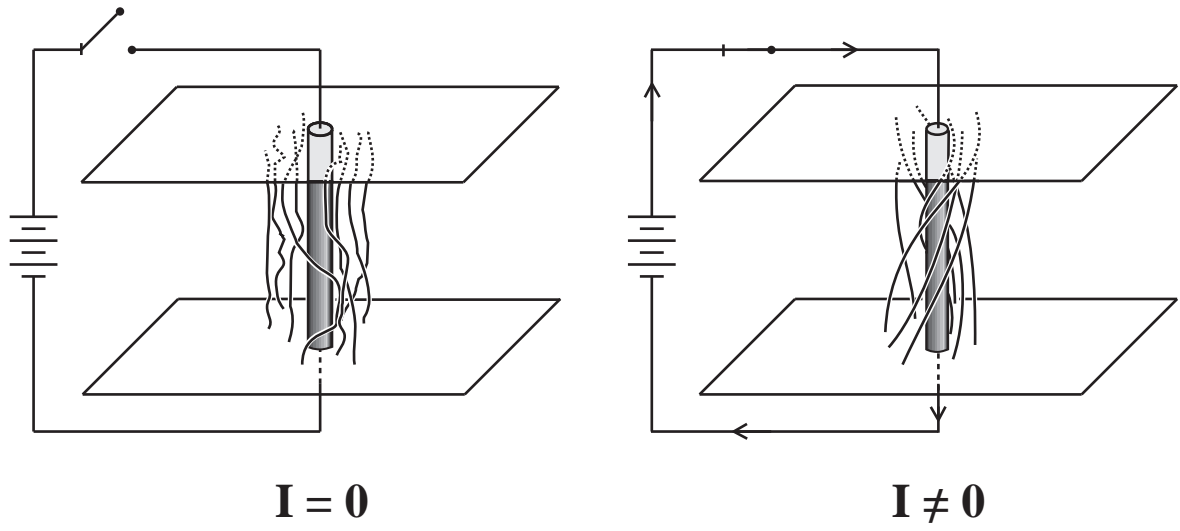
FIG. 2. \ Effect of a current through a cylindrical obstacle 
on a flux liquid. The lines wind around the 
obstacle with a preferred chirality 
when the current is on.

FIG. 3. \  Boson trajectories in $2+1$ dimensions projected 
onto the $xy$-plane. Three trajectories with characteristic
interparticle distance $a$ and four images of the periodic 
box of size $D$ are shown. The periodic boundary conditions
in the time-like direction mean that only small ring 
trajectories appear in the high temperature image at left. 
Trajectories which cross between the periodic cells and lead 
to nonzero winding numbers are shown in the low-temperature 
image on the right.

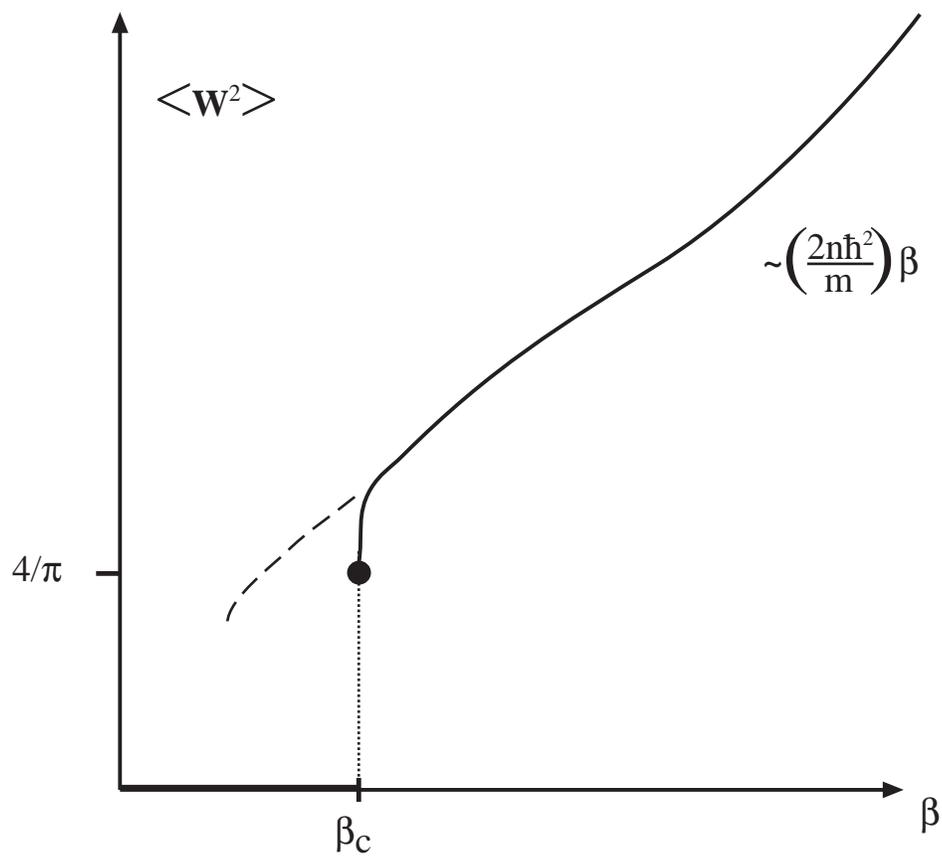
FIG. 4. \  Mean square winding number $\la\W^2\ra$ for bosons
in $2+1$ dimensions a function of $\beta=1/k_BT$. A 
Kosterlitz-Thouless transition occurs for $\beta=\beta_c$. 
Dashed line is the corresponding quantity for directed 
polymer melts.

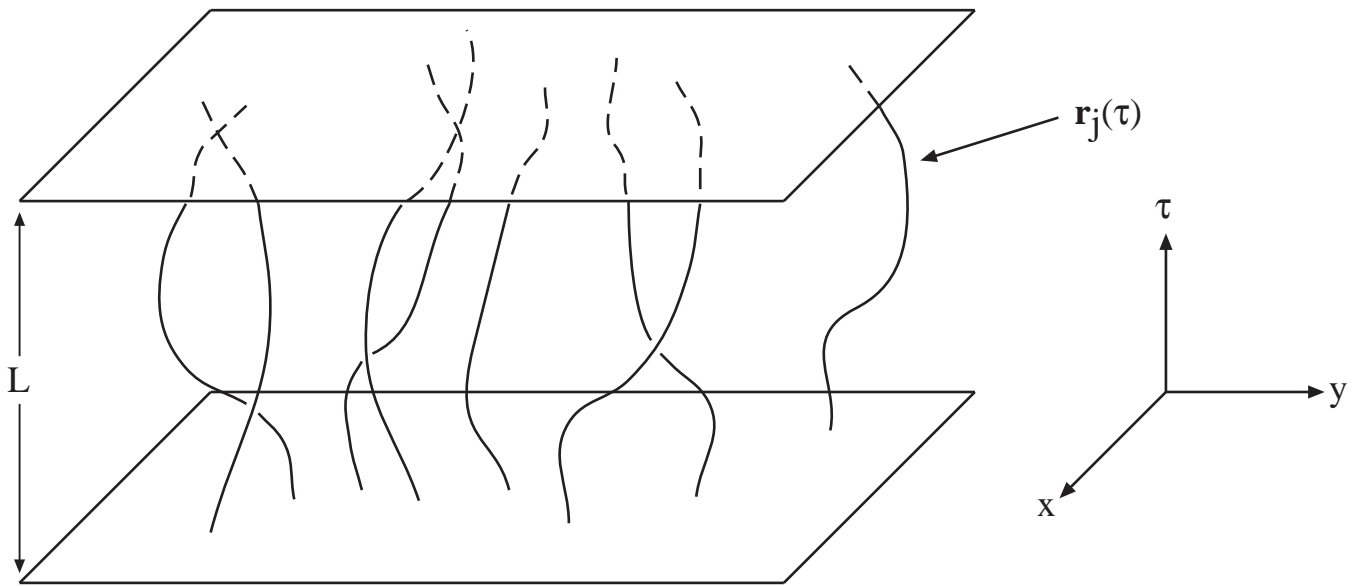
FIG. 5. \  Trajectories for directed polymer melts with 
``open'' boundary conditions at the top and bottom oif a 
slab of thickness $L$.
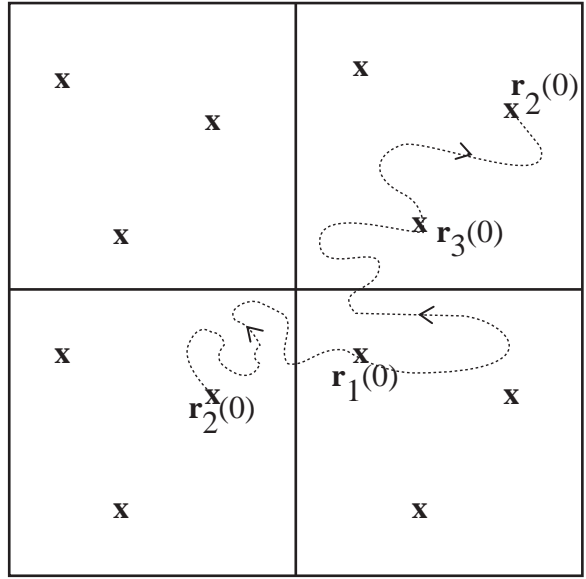
\end{document}